# The suitability of h and g indexes for measuring the research performance of institutions[1]

*Giovanni Abramo*[a,b,*], *Ciriaco Andrea D'Angelo*[b], *Fulvio Viel*[b]

[a] Institute for System Analysis and Computer Science (IASI-CNR)
National Research Council of Italy

[b] Laboratory for Studies of Research and Technology Transfer
School of Engineering, Department of Management
University of Rome "Tor Vergata"

**Abstract**

It is becoming ever more common to use bibliometric indicators to evaluate the performance of research institutions, however there is often a failure to recognize the limits and drawbacks of such indicators. Since performance measurement is aimed at supporting critical decisions by research administrators and policy makers, it is essential to carry out empirical testing of the robustness of the indicators used. In this work we examine the accuracy of the popular "h" and "g" indexes for measuring university research performance by comparing the ranking lists derived from their application to the ranking list from a third indicator that better meets the requirements for robust and reliable assessment of institutional productivity. The test population is all Italian universities in the hard sciences, observed over the period 2001-2005. The analysis quantifies the correlations between the three university rankings (by discipline) and the shifts that occur under changing indicators, to measure the distortion inherent in use of the h and g indexes and their comparative accuracy for assessing institutions.

* **Corresponding author**: Dipartimento di Ingegneria dell'Impresa, Università degli Studi di Roma "Tor Vergata", Via del Politecnico 1, 00133 Rome - ITALY, tel/fax +39 06 72597362, giovanni.abramo@uniroma2.it

# 1. Introduction

The measurement of research performance in higher education institutions is intended to support decisions of policy makers and administrators, and also to assist students, researchers, private companies and other stakeholders in their various choices, by reducing asymmetric information on research quality.

Evaluation of research activity involves complex tasks that should be conducted with maximum methodological rigor, because the results inform critical decision-making processes in the context of the current knowledge economy. However the need for methodological rigor may be seen as conflicting with government and administrative needs for quick and "clear" information. Even among actual evaluation practitioners, simplicity and rapidity can prevail over rigor and reliability. This helps explains the rapid success achieved for the h-indicator, originally suggested by the physicist J.E. Hirsch in 2005. Hirsch's proposal (Hirsch, 2005) attracted rapid and very broad international interest (his original article now counts almost 1,300 citations in SCOPUS and 1,100 on Web of Science), because his indicator represented a single whole number that could quickly summarize both the quantity and impact of a scientist's portfolio of work[2]. The literature is indeed rich of works stressing faults and limitations of this index and warning about its use (Waltman and Van Eck, 2012; Schreiber et al., 2012; Vinkler, 2013). Also limiting the focus to a theoretical perspective, it's ascertained the lack of fulfillment of stability, monotonicity and concavity properties (Ravallion and Wagstaff, 2011).

Nevertheless, evaluation exercises based on the h-index and its variants have proliferated over the years and have often supported important decisions. As just one example, reaching associate or full professor status in Italy requires that the candidate achieve thresholds in three bibliometric indicators, one of which is based on the h-index. Hirsch's proposal found ready users in practical applications, but also attracted great attention among scholars in scientometrics. Certain works took Hirsch's idea, noted the advantages and proposed more or less appropriate applications of the *h-index* to new analytical contexts: journals, research groups, countries, etc. (Braun et al., 2006; Van Raan, 2006; Vanclay, 2008; Guan and Gao, 2008). Others concentrated on the predictive power of the indicator and attempted to validate its robustness, for application in place of more complex and better known indicators (Hirsch, 2007; Hönekopp and Klebe, 2008; Jensen et al., 2009; Rezek et al., 2011; Carbon, 2011; Hönekopp and Khan, 2012).

There has recently also been a significant body of literature suggesting use of the h-index for analysis at the organizational level. Sombatsompop et al. (2011) propose the evaluation of scientific performance in 12 Asian universities, active in fuel and energy research, based on the h-index, aggregated with other standard bibliometric indicators. Similarly, Lazaridis (2010) proposes the h-index for assessing individual professors in chemistry, chemical engineering, materials science, and physics, but then mean values for ranking their entire departments. Again, Franceschini and Maisano (2011) propose a comparison of 33 academic research groups within one discipline (production technology and manufacturing systems), by a structured method using h-based bibliometric indicators.

---

[2] The "*h-index*" represents the maximum number *h* of works by a scientist that have at least *h* citations each.

The specific use of the h-index to rank institutions has been the subject of notable attention by certain scholars, who have investigated potential distortions inherent in Hirsch's original formulation. At the theoretical level, Marchant and Bouyssou (2011) discuss the properties of "consistency" and independence of the h-index, and warn against using a single method to compute both h rankings for scientists and for entire departments. Molinari and Molinari (2008) express concerns over its size dependence, and propose a complement to the h-index for use in comparing the scientific production of institutions (universities, laboratories or journals) with research staff of different numerosity. Huang and Lin (2012) recently compared three different methods for counting publications (whole, straight, and fractional counting) when ranking universities by the h-index. They use a large bibliometric dataset composed of physics papers indexed in Web of Sciences (WoS) over 20 years and ascribed to 299 universities, sorted based on h-index scores. They show that the three counting approaches resulted in observable differences in the h-index scores and institution ranks. Lastly, Kuan et al., 2012 analyze 300 worldwide institutions with publications between 2008 and 2009 in 40 subject categories of clinical medicine. They demonstrate that the h-index compresses the differences among institutions because it ignores the true impact of the works in the h-core, indicating organizations that have very different total impact as being equally performing.

Hirsch's ingenious concept of a single indicator that measures both quantity and quality of research production is one that goes in the right direction. The h-index can in fact be considered the embryo of a true measure of productivity, in that it identifies the desired outcome of research activity as the actual value of scientific advancement (approximated by citations), rather than as being simply "publication". However taking this very view, various scholars have identified a range of limits of the h-index as measure and have proposed a number of variants. To begin, the h-index inevitably ignores the impact of works with a number of citations below h and all the citations above h of the "h-core" works, which is often a very consistent share (Ye and Rousseau, 2010; and Zhang, 2009). The g-index[3] was conceived to take account of the citations above h but did not entirely solve the limits, because it still neglects all citations outside the g-core works. In measuring impact it is also necessary to consider the specific field (subject category) for each of the scientist's publications and carry out appropriate field-normalization. For this purpose, Radicchi et al. (2008) proposed a "generalized h-index", which rescales the number of citations by the average of their distribution in the paper's field. In measuring productivity one should also account for the number of co-authors and their position in the byline, where this is meaningful. Thus Batista et al. (2006) proposed dividing the h-index of a researcher by the average number of authors in the papers considered. Last but not least, because of the different intensity of publications across fields, productivity rankings need to be carried out by field (Abramo and D'Angelo, 2007), when in reality there is a strong interest in comparing researchers from different fields and a human weakness to do just that, through the problematic h-indexes. Iglesias and Pecharromán (2007) tried to correct this flaw by introducing a multiplicative correction which depends basically on the (WoS) field the author is in, and to some extent, on the number of papers the researcher has published. Each h-variant indicator tackles one of the many drawbacks of the h-index while leaving the others unsolved, so none can be considered completely satisfactory. In a previous work we proposed a proxy measure of individual researcher productivity that better meets all the necessary requirements (Abramo and D'Angelo, 2011). We called it "fractional scientific strength" (FSS). In a recent work, taking the FSS as benchmark, we assessed the level of distortions in the rankings of individual scientists' research performance by both the h and g indexes (Abramo et al., 2012a).

---

[3] The g-index represents the highest number "g" of articles that together received $g^2$ or more citations (Egghe, 2006).

This work continues from the earlier one, again comparing the distortions from the h and g indexes but this time at the organizational level, the university. There are two objectives: to quantify the levels of accuracy for the *h* and *g* indexes for evaluating research performance at the level of the research organizations and to obtain practical information on whether the *g-index* represents any true improvement over the *h-index* in measuring such performance. The findings from this current analysis will be compared against the previous results to determine if distortion increases or diminishes in proceeding from the individual to the organizational level. The field of observation is all Italian universities in the hard sciences over the period 2001-2005.

Section 2 presents the main characteristics of the dataset and the three indicators used in the analysis. Section 3 provides the outcomes of comparisons of ranking lists built using the indicators, with a further in-depth analysis concerning the specific subset of universities that place at the top of the rankings. The final section summarizes the results of the work, compares them to previous assertions in the literature, and discusses their implications.

## 2. Methodology

### 2.1 Dataset

The bibliometric dataset used in the analysis is extracted from the Italian Observatory of Public Research (ORP), a database developed and maintained by the authors and derived under license from the WoS. Beginning from the raw data of publications with author address in Italy, and applying a complex algorithm for reconciliation of the author's affiliation and disambiguation of the true identity of the authors, each publication is attributed to the academic scientists that produced it (D'Angelo et al., 2011).

The proposed analysis is based on publications (articles, reviews and conference proceedings only) authored by Italian academic scientists in the period 2001-2005. Citations are observed as of 30/06/2009, providing a sufficient window to guarantee a reliable impact assessment (Abramo et al., 2012b). We take advantage of a unique feature of the Italian university system, in which each academic is classified in one and only one scientific field. In the hard sciences there are 205 such fields (named scientific disciplinary sectors, SDSs), grouped into nine disciplines (named university disciplinary areas, UDAs[4]). To assure full representativeness of publications as proxy of the research output, the field of observation is limited to those SDSs (184 in all[5]) where at least 50% of researchers produced at least one publication in the observed period.

The identification of the research staff and their SDS classifications, for each university, is accomplished by referring to a database on all Italian personnel maintained by the Ministry of Universities and Research[6]. In the five years under examination and the 184 SDSs considered, there were over 37,000 scientists (assistant, associate and full professors) on staff at least for one year, working in a total of 68 universities. Their distribution by UDA is shown in Table 1. They authored a total of over 140,000 publications, receiving over 1.7 million citations by 30/06/2009.

---

[4] Mathematics and computer sciences; physics; chemistry; earth sciences; biology; medicine; agricultural and veterinary sciences; civil engineering; industrial and information engineering.
[5] The list is accessible on http://www.disp.uniroma2.it/laboratorioRTT/TESTI/Indicators/ssd2.html

*Table 1: Number of Italian universities, research staff, SDSs, publications and citations per UDA: data for 2001-2005*

| UDA | N. of SDSs | N. of universities | Research staff | Publications | Citations | Cit. per public. |
|---|---|---|---|---|---|---|
| Mathematics and computer science | 9 | 59 | 3,230 | 11,504 | 58,575 | 5,1 |
| Physics | 8 | 59 | 2,738 | 21,737 | 271,473 | 12,5 |
| Chemistry | 12 | 58 | 3,449 | 22,570 | 304,619 | 13,5 |
| Earth sciences | 12 | 48 | 1,407 | 3,815 | 35,909 | 9,4 |
| Biology | 19 | 63 | 5,423 | 24,719 | 411,131 | 16,6 |
| Medicine | 47 | 55 | 11,803 | 42,103 | 699,641 | 16,6 |
| Agricultural and veterinary science | 28 | 42 | 2,915 | 7,615 | 71,682 | 9,4 |
| Civil engineering | 7 | 45 | 1,338 | 3,261 | 18,357 | 5,6 |
| Industrial and information engineering | 42 | 61 | 4,899 | 25,181 | 145,811 | 5,8 |
| Total | 184 | 68 | 37,202 | 142,431* | 1,731,900* | 12,2 |

*\* These values differ from the column totals due to multiple counts for publications by co-authors belonging to different UDAs.*

## 2.2 Indicators

Our productivity indicator FSS is based on the following reasoning. Research activity is a production process whose output is new knowledge. The principal efficiency indicator of any production system is labor productivity, i.e. the ratio of the value of output to input. When measuring labor productivity, if there are differences in the production factors (capital, scientific instruments, materials, etc.) available to each scientist then one should normalize by them. Unfortunately, in Italy relevant data are not available at individual level. We assume then that resources available to researchers within the same field of observation are the same. A further assumption is that the hours devoted to research are more or less the same for all researchers. These assumptions are fairly well satisfied in the Italian higher education system, which is mostly public and not competitive. Up to 2009, the core funding by government was input oriented, meaning that it was distributed to universities in a manner intended to satisfy the needs for resources of each and all, in function of their size and activities. Furthermore, the time to devote to education is established by law.

In the hard sciences most of the new knowledge produced is codified in publications indexed by such bibliometric databases as WoS or Scopus. As proxy of the value of output we adopt the number of citations for the researcher's publications. Because the intensity of publications varies by field, we compare researchers within the same field, meaning the same SDS. It is very possible though that researchers belonging to a particular scientific field will also publish outside that field. Because citation behavior varies by field, we standardize the citations for each publication with respect to the median of the distribution of citations for all the Italian cited-only publications[7] of the same year and the same WoS subject category. Because research projects frequently involve a team of researchers, which shows in co-authorship of publications, we account for both the fractional contributions of scientists to outputs, as the reciprocal of number of co-authors, and their position in the byline.

---

[6] http://cercauniversita.cineca.it/php5/docenti/cerca.php, last accessed on September 30, 2012.

To quantify the levels of accuracy for the *h* and *g* indexes, we measure the performance of each Italian university in the dataset in the SDS *s* by three different indicators:

$$P_s(h) = \frac{h_s}{RS_s}$$

$$P_s(g) = \frac{g_s}{RS_s}$$

$$P_s(FSS) = \frac{1}{RS_s} \cdot \sum_{i=1}^{N_s} \frac{c_i}{Me_i} \cdot f_{i,s}$$

With:
$RS_s$ = full time equivalent research staff of university in SDS *s*, in the observed period;
$h_s$ = *h index* of the scientific portfolio[8] of all researchers of university in SDS *s*
$g_s$ = *g index* of the scientific portfolio[9] of researchers of university in SDS *s*
$N_s$ = total number of publications of university in SDS *s*;
$c_i$ = citations received by publication *i*;
$Me_i$ = median of the distribution of citations received for all Italian cited publications of the same year and subject category of publication *i*;
$f_{i,s}$ = fractional contribution of authors of publication *i* of university in SDS *s*.

In the life sciences, widespread practice is to indicate the various individual contributions to published research by the positioning of the names in the authors byline. For the life sciences then, when the number of co-authors is above two, *f* is computed giving different credits to each co-author according to their position in the byline and the character of the co-authorship (intra-mural or extra-mural). If first and last authors belong to the same university, 40% of citations are attributed to each of them; the remaining 20% are divided among all other authors. If the first two and last two authors belong to different universities, 30% of citations are attributed to first and last authors; 15% of citations are attributed to second and last author but one; the remaining 10% are divided among all others[10].

Next we compare the ranking lists resulting from the above indicators, for each of the 184 SDSs under observation. We will consider the P(FSS) ranking as the benchmark since is really based on a productivity index i.e., the principal efficiency indicator of any production system. So, we will quantify the levels of accuracy for the *h* and *g* indexes by comparing the P(h) and P(g) rankings to the benchmark.

---

[7] As frequently observed in literature (Lundberg, 2007), standardization of citations with respect to median value rather than to the average is justified by the fact that distribution of citations is highly skewed in almost all disciplines.
[8] A publication co-authored by researchers of the same SDS and university is considered only once.
[9] See note 5.
[10] These percentages for weighting were assigned following the results of interviews of top Italian professors in the life sciences: the values could be changed to suit practices in other national contexts.

## 3. Results

### 3.1 Correlation analysis of rankings

First we present the case of a single SDS, then extend the analysis to all SDSs and finally aggregate data for entire UDAs. Table 2 shows the case of CHIM/07-Foundations of Chemistry for Technologies, in the Chemistry UDA. Columns 2, 3 and 4 show the absolute value of P(FSS), P(h) and P(g) for each of the 35 universities with a research staff in this SDS. The universities are ordered according to their ranking by P(FSS) value. The correlation index between P(FSS) and P(h) is $0.52^{11}$, only slightly lower than that between P(FSS) and P(g), at $0.55^{10}$. We also observe an almost perfect correlation between P(h) and P(g), with the correlation index being 0.97.

*Table 2: Values of P(FSS), P(h) and P(g) for universities active in CHIM/07, ordered by decreasing P(FSS); (ranks for P(h) and P(g) in parentheses)*

| University | P(FSS) | P(h) | P(g) |
|---|---|---|---|
| ID1 | 9.65 | 1.64 (1) | 2.21 (2) |
| ID2 | 4.67 | 0.74 (2) | 1.18 (11) |
| ID3 | 3.87 | 0.86 (7) | 1.21 (10) |
| ID4 | 3.66 | 0.90 (6) | 1.60 (3) |
| ID5 | 3.30 | 1.00 (3) | 1.42 (4) |
| … | … | … | … |
| ID16 | 1.72 | 1.38 (2) | 2.63 (1) |
| ID17 | 1.71 | 0.58 (20) | 0.95 (17) |
| ID18 | 1.53 | 0.30 (29) | 0.50 (28) |
| ID19 | 1.37 | 0.27 (34) | 0.34 (35) |
| ID20 | 1.36 | 0.20 (35) | 0.35 (34) |
| … | … | … | … |
| ID31 | 0.57 | 0.75 (12) | 1.13 (12) |
| ID32 | 0.53 | 0.80 (10) | 1.40 (6) |
| ID33 | 0.45 | 0.39 (25) | 0.52 (26) |
| ID34 | 0.42 | 0.80 (11) | 1.00 (14) |
| ID35 | 0.40 | 0.40 (24) | 0.70 (24) |

We extend the analysis to all the SDSs: Figure 1 presents the distribution of Spearman correlation indexes for rankings based on P(FSS) on the one hand, and P(h) or P(g) on the other hand. For reasons of significance, we limit the analysis to the 164 SDSs with at least 10 universities having active research staff over the five years examined. The two curves show a similar trend, but with the exception of the few SDSs where correlation is virtually inexistent (right tail), the curve for the P(FSS) to P(g) comparison is always above the P(FSS) to P(h) curve. In the P(FSS) to P(h) comparison, there are 45 SDSs (27% of total) with correlation greater than 0.8, while 112 (about 68% of total) show correlation greater than 0.6 and 25 (about 15% of total) show correlation less than 0.4. In slight contrast, 69 SDSs (42%) show P(FSS) to P(g) correlation greater than 0.8, and 122 SDSs (74%) have correlation higher than 0.6, while 16 SDSs (10%) show correlation lower than 0.4. In BIO/10 (Biochemistry), and in INF/01 (Computer Science) we observe the lowest correlations: 0.04 for P(FSS) to P(h) and 0.01 for P(FSS) to P(g), even if both values are not meaningful (*p-value* > 0.7). Indeed, low correlation values are often associated with high *p-values*. In Table 3 we present average values of Spearman correlation for those SDSs only (130 out of 164)

---

[11] *p-value* < 0.01

where correlation analysis is meaningful, i.e. *p-value* < 0.01, aggregated by UDA. Spearman correlation is very strong for rankings by P(h) and P(g) with a low dispersion around the average value (0.94). However the correlation values for the comparisons to P(FSS) ranking are lower. For P(FSS) to P(h) rankings, the correlation shows a minimum of 0.64 for Chemistry and Mathematics, a maximum of 0.82 for Earth sciences and Agricultural and veterinary science, and an overall average value of 0.74. Correlation between P(FSS) and P(g) is quite similar: minimums and maximums are in Chemistry (0.64) and Agricultural and veterinary science (0.84), with an overall average of 0.76.

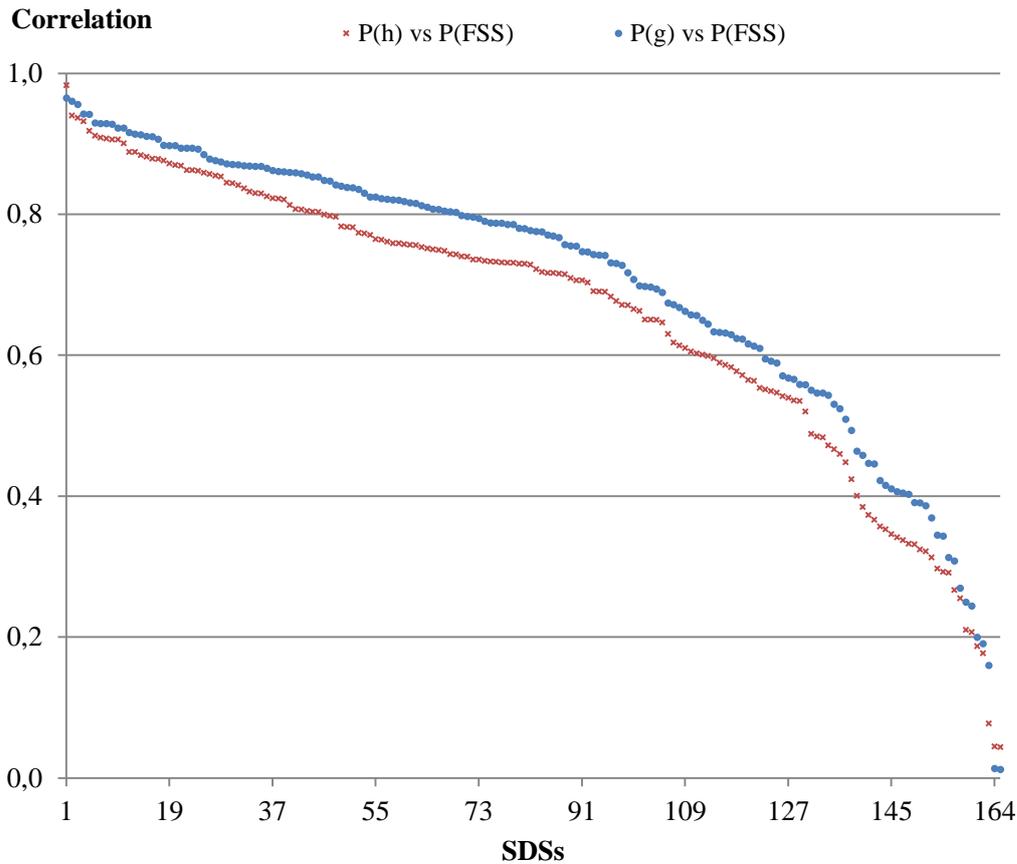

*Figure 1: Spearman correlation for rankings based on P(h), P(g) and P(FSS): distribution by SDSs*

*Table 3: Spearman correlation for rankings based on P(h), P(g) and P(FSS): average values for SDSs with meaningful correlation (p-value < 0.01)*

| UDA | SDS | Spearman correlation | | |
|---|---|---|---|---|
| | | P(FSS) vs P(h) | P(FSS) vs P(g) | P(h) vs P(g) |
| Mathematics and computer science | 8 (out of 9) | 0,64 | 0,70 | 0.93 |
| Physics | 3 (out of 8) | 0,78 | 0,66 | 0.95 |
| Chemistry | 4 (out of 11) | 0,64 | 0,64 | 0.95 |
| Earth sciences | 8 (out of 11) | 0,82 | 0,78 | 0.95 |
| Biology | 16 (out of 19) | 0,68 | 0,71 | 0.94 |
| Medicine | 38 (out of 42) | 0,73 | 0,76 | 0.95 |
| Agricultural and veterinary science | 21 (out of 28) | 0,82 | 0,84 | 0.93 |
| Civil engineering | 7 (out of 7) | 0,79 | 0,82 | 0.94 |
| Industrial and information engineering | 25 (out of 29) | 0,73 | 0,78 | 0.95 |
| Total/average | 130 (out of 164) | 0,74 | 0,76 | 0.94 |

## 3.2 Analysis of quartile variations in rankings

In the section above we saw that ranking lists derived from indicators based on the h and g indicators certainly show a significant correlation to the ranking by FSS. However an overall positive correlation does not preclude the existence of notable shifts in rank among single universities, which could have serious consequences in the actual use of the ranking lists. In this section we further compare the lists by dividing them in quartiles and analyzing the shifts detected when we use P(h) or P(g) instead of P(FSS) for measuring university research performance.

Table 4 shows quartile performance registered under the three indexes (Columns 2, 3, 4) for examples of fifteen universities active in CHIM/07. The ID numbers are assigned according to P(FSS) rank. The last two columns show the absolute value of the quartile shift between the P(FSS) ranking and the P(h) or P(g) rankings.

*Table 4: Quartiles and quartile variations of performance measured by P(FSS), P(h) and P(g), for universities active in SDS CHIM/07*

| University | Quartile | | | Quartile variations | |
|---|---|---|---|---|---|
| | P(FSS) | P(h) | P(g) | P(FSS) vs P(h) | P(FSS) vs P(g) |
| ID1 | 1 | 1 | 1 | 0 | 0 |
| ID2 | 1 | 2 | 2 | 1 | 1 |
| ID3 | 1 | 1 | 2 | 0 | 1 |
| ID4 | 1 | 1 | 1 | 0 | 0 |
| ID5 | 1 | 1 | 1 | 0 | 0 |
| … | … | … | … | … | … |
| ID16 | 2 | 1 | 1 | 1 | 1 |
| ID17 | 2 | 3 | 2 | 1 | 0 |
| ID18 | 2 | 4 | 4 | 2 | 2 |
| ID19 | 3 | 4 | 4 | 1 | 1 |
| ID20 | 3 | 4 | 4 | 1 | 1 |
| … | … | … | … | … | … |
| ID31 | 4 | 2 | 2 | 2 | 2 |
| ID32 | 4 | 2 | 1 | 2 | 3 |
| ID33 | 4 | 3 | 3 | 1 | 1 |
| ID34 | 4 | 2 | 2 | 2 | 2 |
| ID35 | 4 | 3 | 3 | 1 | 1 |
| | | | Total | 28 | 28 |

In this SDS, out of an overall 35 active universities, there are 22 that register shifts in quartile between the P(FSS) ranking and both the P(h) and P(g) rankings. The average value of quartile shift comparing to both P(h) and to P(g) is 0.8. The table is presented solely as an example of this type of analysis, here including four cases of universities registering two-quartile shifts in changing from P(FSS) to P(h). In the comparison between the ranking by P(FSS) and by P(g) there is even a case of a university (ID32) that is in the last quartile for P(FSS) but in the first quartile for P(g).

Table 5 provides a synthesis of the events for all SDSs in the Physics UDA. In the smallest SDS (FIS/08 - Didactics and history of physics) less than a third of the institutions change quartile class when ranked by P(h) (20%) or P(g) (27%). In almost all other cases, at least 50% of universities change quartile. The largest SDS (FIS/01- Experimental Physics) shows the highest percentage of quartile shifts in comparing P(FSS) rankings to P(h) or P(g) rankings. In this SDS, three quarters of universities change quartile when ranked with indicators different from P(FSS), with an average quartile shift equal to 1.15 in the P(FSS) to P(h) comparison, and 1.19 in the P(FSS) to P(g) comparison. In a full five SDSs there are institutions that shift from the top quartile when ranked by P(FSS) to the bottom when ranked by P(h) or P(g), or vice versa (column 7 and 8).

*Table 5: Statistics of quartile variations of performance measured by P(FSS), P(h) and P(g) for universities active in UDA Physics*

| SDSs | Universities | Universities registering quartile variations (%) | | Average quartile variation | | Max quartile variation | |
|---|---|---|---|---|---|---|---|
| | | P(FSS) vs P(h) | P(FSS) vs P(g) | P(FSS) vs P(h) | P(FSS) vs P(g) | P(FSS) vs P(h) | P(FSS) vs P(g) |
| FIS/01 | 52 | 73 | 77 | 1.15 | 1.19 | 3 | 3 |
| FIS/02 | 36 | 72 | 75 | 1.06 | 1.06 | 3 | 3 |
| FIS/03 | 41 | 63 | 59 | 0.93 | 0.83 | 3 | 3 |
| FIS/04 | 30 | 67 | 70 | 1.07 | 1.13 | 3 | 3 |
| FIS/05 | 24 | 50 | 54 | 0.58 | 0.58 | 2 | 2 |
| FIS/06 | 22 | 55 | 45 | 0.73 | 0.55 | 2 | 2 |
| FIS/07 | 45 | 69 | 69 | 1.07 | 1.02 | 3 | 3 |
| FIS/08 | 15 | 20 | 27 | 0.27 | 0.27 | 2 | 1 |

Table 6 shows the extreme cases for each UDA, specifically the details for the two SDSs with the maximum and the minimum percentages of universities registering quartile variations between P(FSS) and P(h) rankings.

*Table 6: Statistics for quartile variations of performance measured by P(FSS) and P(h): for each UDA, the table shows the two SDSs with the maximum and the minimum percentages of universities registering quartile variations*

| UDAs | SDS | Universities registering quartile variations (%) | Average quartile variation | Max quartile variation |
|---|---|---|---|---|
| Mathematics and computer science | INF/01 | 78 | 1.3 | 3 |
| | MAT/01 | 35 | 0.4 | 1 |
| Physics | FIS/01 | 73 | 1.2 | 3 |
| | FIS/08 | 20 | 0.3 | 2 |
| Chemistry | CHIM/06 | 79 | 1.1 | 3 |
| | CHIM/08 | 45 | 0.6 | 2 |
| Earth sciences | GEO/08 | 73 | 1.2 | 3 |
| | GEO/04 | 31 | 0.4 | 3 |
| Biology | BIO/10 | 85 | 1.4 | 3 |
| | BIO/07 | 33 | 0.5 | 2 |
| Medicine | MED/26 | 69 | 1.0 | 3 |
| | MED/34 | 15 | 0.2 | 1 |
| Agricultural and veterinary science | VET/01 | 69 | 1.0 | 3 |
| | AGR/09 | 18 | 0.2 | 2 |
| Civil engineering | ICAR/08 | 54 | 0.6 | 2 |
| | ICAR/05 | 20 | 0.2 | 1 |
| Industrial and information engineering | ING-INF/03 | 68 | 1.1 | 3 |
| | ING-IND/26 | 27 | 0.3 | 1 |

Table 7 shows the synthesis of data from all the analyses conducted for the SDSs under observation, grouped by UDA. Over all the UDAs, the shifts in quartile involve an average of roughly 50% of the universities evaluated (last line of Columns 2 and 3). The variation in quartile averages 0.7 for the P(FSS) to P(h) comparison and 0.6 for P(FSS) to P(g). For these same comparisons, an average of 14.4% and 12.6% of total universities change at least two quartiles (last line of Columns 6 and 7). The UDA most sensitive to the type of indicator used is Chemistry: 62.5% of the universities change quartile when their performance is measured using P(h) in place of P(FSS), and 20.9% of the universities show a shift of at least two quartiles. On the other hand, in Civil engineering, shifts in quartile between ranking lists for P(FSS) and P(h) concern only 42% of the universities, with the average value of quartile shift at 0.5, while 7.6% of universities experience shifts greater than one quartile. The shifts for the comparison between P(FSS) and P(g), while still notable, are consistently fewer than those for the P(FSS) to P(h) comparison.

*Table 7: Statistics of quartile variations of performance measured by P(FSS), P(h) and P(g), by UDA*

| UDA | Universities registering quartile variations (%) | | Average quartile variation | | Universities registering quartile variations ≥2 (%) | |
|---|---|---|---|---|---|---|
| | P(FSS) vs P(h) | P(FSS) vs P(g) | P(FSS) vs P(h) | P(FSS) vs P(g) | P(FSS) vs P(h) | P(FSS) vs P(g) |
| Mathematics and computer science | 56.2 | 55.2 | 0.8 | 0.7 | 15.9 | 15.7 |
| Physics | 58.6 | 59.5 | 0.9 | 0.8 | 22.5 | 19.7 |
| Chemistry | 62.5 | 61.2 | 0.9 | 0.9 | 20.9 | 20.5 |
| Earth sciences | 48.9 | 43.7 | 0.6 | 0.5 | 12.9 | 10.0 |
| Biology | 52.9 | 50.7 | 0.7 | 0.7 | 16.1 | 12.3 |
| Medicine | 48.3 | 45.3 | 0.6 | 0.6 | 12.2 | 9.7 |
| Agricultural and veterinary science | 45.8 | 44.4 | 0.6 | 0.6 | 10.4 | 9.5 |
| Civil engineering | 42.0 | 40.8 | 0.5 | 0.5 | 7.6 | 6.5 |
| Industrial and information engineer. | 49.0 | 46.1 | 0.6 | 0.6 | 11.5 | 9.4 |
| Total | 51.6 | 49.6 | 0.7 | 0.6 | 14.4 | 12.6 |

Table 8: presents further data on the comparison between rankings derived from the three performance indicators, here showing the number of universities that place above the national median for P(FSS) but below for P(h) (column 2) or P(g) (column 3).

One again, we see a critical situation in Chemistry, where the performance for 20 universities (out of 58 total in the UDA) would drop below the national median if measured by P(h), even though the value of P(FSS) shows they are above median. Immediately after comes the Physics UDA, where 19 out of 59 universities experience the same event. Again, the rankings for P(g) show notable but slightly less serious cause for concern.

*Table 8: Universities with P(FSS) above the national median, not included in the same subset when performance is measured by P(h) or P(g)*

| | Universities above the median by P(FSS) shifting below by | |
|---|---|---|
| UDA | P(h) | P(g) |
| Mathematics and computer science | 16 (out of 59) | 14 (out of 59) |
| Physics | 19 (out of 59) | 17 (out of 59) |
| Chemistry | 20 (out of 58) | 18 (out of 58) |
| Earth sciences | 12 (out of 48) | 9 (out of 48) |
| Biology | 14 (out of 63) | 12 (out of 63) |
| Medicine | 11 (out of 55) | 9 (out of 55) |
| Agricultural and veterinary science | 8 (out of 42) | 9 (out of 42) |
| Civil engineering | 7 (out of 45) | 8 (out of 45) |
| Industrial and information engineering | 12 (out of 61) | 9 (out of 61) |

If h or g indicators were used to measure university performance and resources were then allocated only to those above the national median (as is the practice in some nations), the levels of distortion would certainly have unfortunate consequences.

Since in some nations (such as with the UK research assessment exercise, RAE[12]), the funding system is conceived precisely to allocate the greater part of resources to "excellent" universities, it would interesting to know what happens at the very top of the rankings when indicators such as P(h) or P(g) are used in place of P(FSS). The next section examines this issue.

### 3.3 Analysis of top universities

For every SDS we now identify the universities included in the first quartile of the ranking by P(FSS), then check which of these would not be included in the same quartile under the rankings constructed with P(h) and P(g). We first analyze the example of the SDSs in the Earth science UDA (Table 9). For the particular case of GEO/04 (Paleontology and paleoecology), 38% of the universities at the top for ranking by P(FSS) would no longer be "top" under the indicator P(h), while 12,5% would lose their standing under P(g).

On average in Earth sciences, 42% of the universities that are ranked as excellent by P(FSS) fail to achieve top level in the ranking derived from P(h), and 35% again do not achieve this status under ranking for P(g). The data for GEO/06 (Mineralogy) are particularly notable: 20 of 24 universities (83% of total) active in the SDS place in the first quartile for P(FSS) but not for P(h). An equally critical situation occurs in GEO/10, where 57% of top universities by P(FSS) fail to achieve this status under P(h) and 43% fail under P(g). There is only SDS GEO/07 (Petrology and petrography) with a case of perfect superimposition of first quartiles: here, the excellent universities under P(FSS) are the same under P(g).

*Table 9: Top 25% universities in each SDS of UDA Earth sciences by P(FSS), not included in the same subset when performance is measured by P(h) and P(g)*

|         | Percentage of top 25% universities by P(FSS) not included in same set by | |
|---------|------|------|
| SDSs    | P(h) | P(g) |
| GEO/01  | 38   | 25   |
| GEO/02  | 22   | 22   |
| GEO/03  | 29   | 29   |
| GEO/04  | 38   | 13   |
| GEO/05  | 22   | 11   |
| GEO/06  | 83   | 33   |
| GEO/07  | 29   | 0    |
| GEO/08  | 71   | 29   |
| GEO/09  | 29   | 29   |
| GEO/10  | 57   | 43   |
| GEO/11  | 40   | 20   |
| Total   | 42   | 35   |

We now extend the analysis to all the SDSs, with Table 10 presenting the data aggregated by UDA. On average, the percentage of top 25% universities by P(FSS) that are not included in the same set by P(h) is 42%. Among the individual UDAs, the figures for this data vary between a minimum of 28% for the universities in Civil engineering and a maximum of 49% for Chemistry. Thus Chemistry is again the most problematic UDA, together with Physics (48%) and Mathematics and computer science (45%). The same three UDASs also have the maximum values of shift when comparing the first quartiles for the P(FSS) and P(g) ranking lists, and in a variation of previous patterns, for Physics and Mathematics and computer science the differences between the first quartiles under P(FSS) and P(g) are also greater than those when comparing P(FSS) to P(h). In all the other UDAs, the intersection of the top 25% of universities by P(FSS) and by P(g) is slightly larger than the intersections of these sets under P(FSS) and P(h).

---
[12] http://www.rae.ac.uk/, last accessed on September 30, 2012.

*Table 10: Top universities by P(FSS) that are not included in the same subset when performance is measured by P(h) and P(g)*

| UDA | Percentage of top 25% universities by P(FSS) not included in the same set by | |
|---|---|---|
| | P(h) | P(g) |
| Mathematics and computer science | 45 | 47 |
| Physics | 48 | 51 |
| Chemistry | 49 | 46 |
| Earth sciences | 42 | 35 |
| Biology | 42 | 36 |
| Medicine | 40 | 35 |
| Agricultural and veterinary science | 41 | 33 |
| Civil engineering | 28 | 26 |
| Industrial and information engineering | 40 | 35 |
| Total | 42 | 38 |

## 5. Discussion and conclusions

The evaluation of research institutions' performance based on the h-index is clearly prone to all the inherent weaknesses of this indicator, as has frequently been argued in recent literature. However, since the measurement of performance is intended to support critical decisions by administrators and policy makers, it is essential to provide true empirical testing of the accuracy of the indicators available for potential use, considering that the the principal efficiency indicator of any production system is labor productivity.

In the present work we have provided a specific quantitative measurement of the distortions inherent in use of the h-index and its best-known variant, the g-index. To do this, we adopted a third index as benchmark: fractional scientific strength (FSS). This is an indicator that measures the impact of the entire scientific production of a given research institution, not just that of the most cited publications, and which normalizes citations by field and accounts for number of co-authors contributing to each publication.

Taking these three indicators, we develop and compare the ranking lists for all Italian universities active in the hard sciences over the period 2001-2005. We construct the ranking lists for each of 164 scientific fields, and the 9 overall disciplines in which they are grouped.

For a third of the 164 SDSs examined, the correlation of the other rankings with those from FSS is less than 0.6, and for 10% it is actually lower than 0.4. When we use the h-index in place of FSS, the ranking lists show shifts in quartile involving more than half the universities, and for 14% of the universities these variations are at least two quartiles. Among the disciplines examined, Physics and Chemistry are particularly problematic: their respective SDSs show indexes of correlation between the lists averaging less than 0.5, and very substantial quartile shifts. In five of the eight Physics SDSs there are universities that shift from the top quartile when ranked by FSS to the bottom when ranked by h, or vice versa. This is clearly due to the fact that the Physics area is the one with the greatest intensity of publication and citation, meaning that the use of indicators that ignore the impact of part of the scientific production (the part extraneous to h-core, which is at times substantial) causes still greater distortion than in other UDAs. A further analysis focused on the upper quartile of the ranking lists shows that, out of the top 25% of universities identified by FSS, an average of 40% fail to reach the same subset when ranked by h.

The analysis suggests that, compared to using the h-index, the g-index slightly reduces distortion: the ranking lists prepared with this indicator are on average more correlated with the benchmark relative to those based on h, and the shifts in position are generally less.

Comparing to a preceding work that assessed the distortions involved in ranking individual scientists' research performance by h and g indexes (Abramo et al., 2012), the current analysis shows that the level of distortion is still greater in proceeding to evaluating the institutions. Limiting the evaluation of research performance to the h thresholds of a scientific portfolio introduces important distortions at the level of individual scientists, which become still greater at the level of research institutions. This should put evaluation practitioners on guard over the temptation to adopt simple bibliometric indicators for assessing universities' productivity. While they are easy to understand, measure and communicate, such indicators entail a level of inaccuracy that could well be unacceptable for most of the intended uses and objectives.